\documentclass[twocolumn]{svjour3}          
\usepackage{helvet}
\usepackage{courier}
\usepackage{color,graphicx}
\usepackage{subfigure}
\usepackage{amsmath}
\usepackage{amssymb}

\def\ignore#1{}

\begin{document}

\title{Modeling Perceived Relevance for Tail Queries without Click-Through Data
}


\author{Changsung Kang \and
        Xiaotong Lin \and
        Xuanhui Wang \and
        Yi Chang \and
        Belle Tseng
}


\institute{Changsung Kang, Xuanhui Wang, Yi Chang, Belle Tseng \at
              Yahoo Labs, Sunnyvale CA \\
              \email{\{ckang,xhwang,yichang,belle\}@yahoo-inc.com}           
           \and
           Xiaotong Lin \at
              University of Kansas\\
              \email{cindylin@ku.edu}
}

\date{Received: date / Accepted: date}

\maketitle

\begin{abstract}
Click-through data has been used in various ways in Web search such
as estimating relevance between documents and queries. Since only
search snippets are perceived by users before issuing any clicks,
the relevance induced by clicks are usually called \emph{perceived
relevance} which has proven to be quite useful for Web search.
While there is plenty of click data for popular queries, very
little information is available for unpopular tail ones. These tail
queries take a large portion of the search volume but search
accuracy for these queries is usually unsatisfactory due to data
sparseness such as limited click information. In this paper, we
study the problem of modeling perceived relevance for queries
without click-through data. Instead of relying on users' click
data, we carefully design a set of snippet features and use them to
approximately capture the perceived relevance. We study the
effectiveness of this set of snippet features in two settings: (1)
predicting perceived relevance and (2) enhancing search engine
ranking. Experimental results show that our proposed model is
effective to predict the relative perceived relevance of Web search
results. Furthermore, our proposed snippet features are effective
to improve search accuracy for longer tail queries without
click-through data.
\end{abstract}

\section{Introduction}\label{sec:intro}

Designing effective ranking functions to satisfy all kinds of
information needs of end users is admittedly difficult. A practical
approach used by commercial search engines is to collect all
possible useful signals or features and combine them together using
techniques such as learning to
rank~\cite{BurShaRen05,ZheZhaZha07,CaoXuLiu06}. Beyond the regular
text matching features (e.g., TF-IDF), click-through data has been
studied
extensively~\cite{Joachims02,CraZoeTay08,Agichtein06,shihao2009,CKang:WSDM2011}.
Noticeable usages of click-through data include propagating
semantic information between queries and
documents~\cite{beeferman:00,Li08Query,Xue:CIKM04,YatTib07},
estimating document
relevance~\cite{Joachims02,CraZoeTay08,Dupret:WSDM10,ZhaJon07,Chapelle09},
defining features in learning to
rank~\cite{Agichtein06,shihao2009}, etc. All the existing works
have demonstrated the unique values of click-through data in
improving search engines from many perspectives. A main advantage
of click-through data is that it contains users' \emph{implicit}
perceived relevance feedback.

A well-known challenge in leveraging click-through data is that
click-through information is very noisy and biased by many factors
such as presentation order and appearance
\cite{Joachims05,clarke2007click,Yue:WWW10}. Many studies such as
\cite{CraZoeTay08,DupPiw08,Guo:WWW09,Chapelle09,Joachims05,Hu:WWW2011}
have attempted to address the position bias to extract the
relevance between documents and queries which is hidden in the
clicks. Since only search snippets are perceived by users before
issuing any clicks, the relevance induced by clicks are usually
called \emph{perceived relevance}. In general, when there are
sufficient clicks information for a query, existing approaches can
estimate the perceived relevance reliably which has been proven to
be quite effective to improve Web search.

While there are plenty of click information for popular or head
queries, unfortunately, very little information is available for
tail ones and existing methods either can not be applied to or can
give unreliable estimation for tail queries due to limited click
data. According to a recent study~\cite{Silverstein:1999}, queries
submitted to Web search engines follow a heavy-tailed power-law
distribution. Thus a large fraction of queries are issued very
infrequently, forming the well-known ``long tail''~\cite{longtail}.
Naturally, the useful signals for such tail queries are very scarce
in search logs. As a result, the benefit of click-through data is
mainly for popular head queries and current search engines usually
perform poor for tail queries~\cite{Downey:2007}. A recent
study~\cite{Goel:2010} shows that almost every individual user has
both head and tail requests for Web search. Thus, poor search
results on tail queries can not only make most of users
unsatisfactory for their immediate requests, but also deteriorate
their overall perceptions of a search engine. Interestingly,
\cite{Goel:2010} has also shown that there is a second-order effect
that satisfactory results for tail requests can significantly boost
the head requests due to increased user satisfaction and resulting
repeat patronage. However, search accuracy for these queries is
usually unsatisfactory due to data sparseness. Thus it remains a
challenge to improve search quality for tail queries.

The importance and uniqueness of tail queries has been noticed
recently. Existing works on tail queries mainly focus on aspects
such as query classification~\cite{Broder:SIGIR07}, query
advertisability~\cite{Pandey:2010}, and query
suggestions~\cite{Song:2010}. Surprisingly, there are few works on
directly improving search accuracy for tail queries, which is the
most important aspect of a search engine.
In this paper, we propose a self-reinforcement way for tail
queries. Motivated by the perceived relevance in click-through
data, our main idea is to capture the perceived relevance based on
search result snippets without requiring any click-through data.
Search result snippets are valuable resources for the following
reasons: (1) Search result snippets are highly correlated with
click data and thus the underlying perceived relevance. (2) The
snippets are summaries of the documents which are the most relevant
passages deemed by the snippet generation methods. Passage level
relevance~\cite{Callan:1994} can be modeled by matching queries
with search snippets.

Specifically, we define a set of snippet features whose goal is to
capture the perceived relevance from multiple perspectives,
including language attractiveness, URL attractiveness, and
query-snippet matching attractiveness. All of these features do not
need any user click data and can be computed solely based on
queries and snippets. We study the effectiveness of this set of
snippet features in two settings: (1) predicting perceived
relevance and (2) enhancing search engine ranking. For (1), we
first estimate perceived relevance for queries which have
sufficient clicks using an existing dynamic Bayesian network model.
We then train a machine learning model to predict the estimated
perceived relevance. For (2), however, it is not straightforward to
incorporate these features into a search process since most of the
features can be only computed after the query-dependent snippets
are generated. We thus propose two strategies to leverage these
snippet features. Our first strategy is to combine the predicted
perceived relevance scores with the original ranking scores to
rerank search results. Our second strategy is to expand the
original ranking features by adding the snippet features to learn a
new ranking function. We show that both strategies can be naturally
incorporated into a search process in different application
scenarios.

We evaluate the usefulness of our defined snippet features based on
a large set of queries and snippet features from a commercial
search engine. Experimental results shows that the defined snippet
features can give good prediction of perceived relevance and it can
also improve the search accuracy significantly.

\section{Related Work} \label{sec:related}

The long tail view was first coined in~\cite{longtail} and has been
observed for many diverse applications like e-commerce and Web
search~\cite{Goel:2010}. Our work is more related to the long tail
study in Web search. For example, \cite{Downey:2007} compared head
queries and tail queries in terms of search accuracy and users
search behaviors. \cite{Broder:SIGIR07} proposed robust algorithms
for rare query classification. \cite{Pandey:2010} studied the
advertisability of tail queries in sponsored search and proposed a
word-based approach for online efficient computation.
\cite{Song:2010} studied query suggestions for rare queries but
their approaches still assume that there is click information to
leverage. In contrast, our work is on directly improving the search
accuracy, which is the most important aspect of a search engine,
for tail queries without any click-through data.

In the past, snippets have been used by many different purposes
such as query classification~\cite{Broder:SIGIR07} and measuring
query similarity~\cite{Sahami:WWW06}. In particular, our work is
related to \cite{Agichtein06}. In \cite{Agichtein06}, some snippet
features such as overlap between the words in title and in query
are used, together with user behavior and click-through features.
The main finding of their study is that click features are the most
useful for general queries. In our work, we focus on tail queries
which do not have any click information. We define a more
comprehensive set of snippet features and discuss different
application scenarios to efficiently leverage these snippet
features.

Our work is related to click models and a number of recent studies
has been conducted to analyze click
data~\cite{Joachims02,CraZoeTay08,ZhaJon07,Chapelle09,Dupret:WSDM10,Hu:WWW2011}.
For example, \cite{Joachims02} examined several rule-based methods
to extract the relative preference between a pair of documents from
click data. Recently, all clicks in a search session are modeled
together and thus the dependency among clicks in different
positions can be modeled. For example, cascade
model~\cite{CraZoeTay08} assume that user sequentially examine
results and stop as soon as a relevant document is clicked.
\cite{Chapelle09} and \cite{Guo:WWW09} analyze click data based on
different Baysian generative models and perceived relevance is
estimated by fitting the models to observed click data.
\cite{Hu:WWW2011} further extends these models to consider intent
diversity. A recent approach~\cite{Dupret:WSDM10} uses a session
utility model to estimate the ``intrinsic relevance'' of each
clicked document. Both~\cite{Chapelle09} and \cite{Dupret:WSDM10}
argued the difference between ``perceived relevance'' and ``true
relevance.'' The main resources they relied on to estimate true
relevance of a click are the session activities \emph{after} the
click. Usually a document which is clicked last is given a higher
relevance score. In our work, we choose to model perceived
relevance since there are no actual click information for tail
queries and it is hard to model what happened afterwards.
Furthermore, all these works only leverage the click information
and have not considered search result snippets.

Our work is also related to click prediction
works~\cite{Agarwal:KDD07,Richardson07,DupPiw08}.
\cite{Agarwal:KDD07} used an existing hierarchy to propagate clicks
to rare events. \cite{DupPiw08} used past clicks to predict future.
\cite{Richardson07} proposed a feature-based method of predicting
the click-through rate for new ads. To the best of our knowledge,
few works have been conducted to predict click-based perceived
relevance for tail queries in Web search. Furthermore, compared
with \cite{Richardson07} which only uses query-dependent features,
we explore a more compressive feature set with both query-dependent
and query-independent features.

\section{Perceived Relevance and User Clicks}\label{sec:click}

Click-through data has been extensively studied recently
\cite{Joachims02,CraZoeTay08,Chapelle09,Hu:WWW2011}. A common
observation is that click data contains users' perceived relevance
feedback and this information is quite effective to improve Web
search. However, click data is noisy and biased by many factors
such as presentation order and appearance
\cite{Joachims05,clarke2007click,Yue:WWW10}.
Many studies such as
\cite{CraZoeTay08,Guo:WWW09,Chapelle09,Joachims05} have attempted
to address the position bias to extract the relevance between
documents and queries which is hidden in the clicks. Technically,
perceived relevance is usually captured by the click probability
given the corresponding search result has been examined by end
users. By definition, perceived relevance is independent of
position bias. In the following, we give a brief review of the
Dynamic Bayesian Network (DBN) model which can effectively extract
the perceived relevance from a click session~\cite{Chapelle09}.

The DBN model is based on the cascade model proposed
in~\cite{CraZoeTay08}. The cascade model assumes that a user
examines the search results sequentially from top to bottom and
decides whether to click a search result. After a document $u$ is
examined, it is either clicked with probability $a_u$ or skipped
with probability ($1-a_u$) where $a_u$ denotes the degree of
attractiveness or perceived relevance. The cascade model assumes
that a user who clicks never comes back and a user who skips always
continue. A click on the $i$-th document means that the user skips
all the documents ranked above and the user is satisfied by the
$i$-th document
$$P(C_i = 1) = a_i \prod_{u = 1}^{i-1}(1-a_u)$$
All above assumptions clearly oversimplify the problem. The model
suffers indeed from only being able to consider sessions with
exactly one click.

\cite{Chapelle09} extends the cascade model and proposes a Dynamic
Bayesian Network (DBN) model to simultaneously model the relevance
of all documents in the search results. The DBN model introduces
the notion of {\em satisfaction} to separately model the relevance
of the landing page and perceived relevance on the search results
page ({\em attractiveness}). Formally, we use binary random
variables $E_i, A_i, S_i$ and $C_i$ to denote examination,
attractiveness, satisfaction, and click of $i$-th document. A
session is generated by the following procedure, assuming $E_1=1$
and all other default values are 0:
\begin{itemize}
  \item For each position $i$, sample an attractiveness
      probability $a_i$ from a Beta prior distribution.
  \item For each position $i$, sample a satisfaction
      probability $s_i$ from a Beta prior distribution.
  \item Repeat for each position $i$
  \begin{itemize}
    \item Sample $A_i=1$ with probability $a_i$. Set $C_i =
        A_i$ if $E_i = 1$
    \item If $C_i= 1$, sample $S_i=1$ with probability
        $s_i$. Otherwise set $S_i = 0$
    \item If $S_i = 1$, set $E_{i+1} = 0$. Otherwise,
        sample $E_{i+1} = 1$ with probability $\gamma$.
  \end{itemize}
\end{itemize}
The parameter $\gamma$ is the perseverance parameter and a user may
give up the search with probability $1-\gamma$ before satisfied.
Assuming users always examine the first position, attractiveness,
indeed is a prediction of the CTR at position 1.
Given a set of click sessions in search logs, we can find a maximum
a posterior (MAP) estimation of $a_i$ and $s_i$ by an EM
algorithm~\cite{Chapelle09}. The obtained $a_i$ denotes the degree
of attractiveness of $i$-th document.

In general, when there are sufficient click information for a
query, existing approaches such as DBN can estimate the perceived
relevance reliably. However, the existing click-based methods only
rely on the click information in search logs but totally ignore the
search snippets. While there are plenty of click information for
popular or head queries, unfortunately, very little information is
available for tail ones and existing methods either can not be
applied to or can give unreliable estimation for tail queries due
to scarce click data. In the next section, we describe our approach
to capturing perceived relevance using search snippets.

\section{Capture Perceived Relevance for Tail Queries}\label{sec:snippet}

Tail queries pose a big challenge to leverage the click-based
methods. Since the ultimate goal of the click-based methods is to
capture the perceived relevance and search results snippets are the
main information sources before a user issues a click, we thus try
to capture perceived relevance for tail queries based on search
snippets in this section.

\subsection{A Motivating Experiment}\label{sec:motivation}
Our hypothesis is that there is a strong correlation between
perceived relevance and snippets in a Web search result page. We
test this hypothesis using a simple experiment as follows. In this
experiment, we collected a set of tuples $(q,u_1,u_2)$ where $u_1$
and $u_2$ are any two URLs that appear in the same search result
page for the query $q$ during some period of time. We then computed
the number of missing query tokens in titles. We examined how
likely $u_1$ is clicked more frequently than $u_2$ when $u_1$ has
fewer missing query tokens than $u_2$. In other words, we want to
estimate the probability
\begin{align*}
P_0&=\mathrm{Prob}(u_1\textrm{ is clicked more frequently than }u_2\\
& |~ \textrm{miss}(t_{u_1})<\textrm{miss}(t_{u_2}))
\end{align*}
where $\textrm{miss}(t_{u_i})$ is the number of missing query
tokens in the title $t_{u_i}$ for $u_i$. We balanced our samples to
eliminate potential position bias by ensuring that $u_1$ is
presented higher than $u_2$ in half of our examples. The estimation
of $P_0$ in our data was 0.74, which is much larger than 0.5 and
shows positive correlation. Furthermore, we observed a stronger
click preference if the title matching has larger difference for
the two URLs.
\begin{align*}
P_1&=\mathrm{Prob}(u_1\textrm{ is clicked more frequently than }u_2\\
& |~ \textrm{miss}(t_{u_1})+1<\textrm{miss}(t_{u_2})) = 0.83.
\end{align*}
This result demonstrates that the snippets with more missing query
tokens in their titles tend to receive fewer clicks than those with
fewer missing query tokens. This makes sense intuitively since a
page with title matched well with queries is more likely to be more
relevant. Note that the title matching is only a single feature
among many possible signals that may influence user clicks. To
model the click behaviors more accurately, we need to seek a more
comprehensive set of such snippet features to capture the perceived
relevance more precisely.

\subsection{Search Snippet Features}\label{sec:features}
Our goal is to develop a comprehensive set of snippet features that
capture the attractiveness of results. We define our features from
the following perspectives: language attractiveness, URL
attractiveness, and query-snippet matching attractiveness. All the
features are summarized in Table~\ref{tab:snippet-features} and we
describe them separately in the following.

\begin{table}\centering
\begin{tabular}{|l|p{5cm}|}
\hline
\multicolumn{2}{|c|}{\textbf{Language Attractiveness Features}}\\\hline
\multicolumn{2}{|l|}{Readability Features}\\\hline
NumChars & Number of characters in snippet\\\hline
NumWords & Number of words in snippet\\\hline
NumSegments & Number of period/ellipsis-separated segments\\\hline
NumWordInitCap & Number of words with initial capitals in snippet\\\hline
FracWordInitCap & Fraction of words with initial capitals in snippet\\\hline
NumCapChar & Number of capital characters in title or URL\\\hline
FracCapChar & Fraction capital characters in title or abstract\\\hline
\multicolumn{2}{|l|}{Word-level Attractiveness}\\\hline
FracAttrWord & Fraction of attractive words\\
\hline
\hline
\multicolumn{2}{|c|}{\textbf{URL Attractiveness Features}}\\
\hline
NumChars & Number of characters in URL\\\hline
TopLevelDomain & The top level domain of URL\\\hline
NumLevelDomain & Number of levels in domain\\\hline
NumViews & Number of views (impressions) of URL\\
\hline
\hline
\multicolumn{2}{|c|}{\textbf{Matching Attractiveness Features}}\\
\hline
NumMatch & Number of all matches in snippet\\\hline
NumUniqMatch & Number of unique matches in snippet\\\hline
NumApxMatch & Number of approximate matches in snippet \\\hline
FracMatch & Fraction of matches in snippet\\\hline
FracApxMatch & Fraction of approximate matches in snippet\\\hline
NumBefMatch & Number of words before the first match\\\hline
NumBtwMatch & Number of extra words between matches\\\hline
IsExactMatch &  Is whole query string exactly matched\\\hline
IsOrderMatch & Are matches in the exact order\\\hline
IsSegMatch & Are all matches occur in a single segment\\\hline
\end{tabular}
\caption{Summary of the snippet features.}
\label{tab:snippet-features}
\end{table}

\subsubsection{Language Attractiveness.} We model the language
attractiveness by two sets of features: readability and word-level
attractiveness. (1) Recent studies such as \cite{clarke2007click}
show that the readability of snippets in a search result page can
directly
impact users' click-through behavior. 
In this work, we define some readability features similar to those
proposed in \cite{CutGua07,KanOrr09,RosOrr07} and also some new
features based on our intuitive judgments and experiments. This set
of features are mainly to model the \emph{syntactic} information of
titles and abstracts of the snippets. For example, the feature
NumSegment measures the number of fragments separated by an
ellipsis or a period in abstracts which in some sense reflects how
easy the snippets can be read. (2) The word-level attractiveness is
to model the language in a \emph{semantic} level. Previous
researches~\cite{clarke2007click,KanOrr09,CutGua07,Richardson07}
also show that some terms in titles (e.g., ``official'' or
``gallery'') specify a certain genre and influence user clicks
noticeably. To identify these words, we use a $t$-test based on the
URL attractiveness values estimated by the DBN model. Specifically,
given head queries with attractiveness of URLs estimated by DBN, we
form two sets of titles, $A$ and $U$, where $A$ includes the titles
of the two most attractive URLs and $U$ includes the titles of the
two most unattractive URLs of every query. An attractive words will
have higher discriminative power between $A$ and $U$ and a less
attractive word will have smaller difference between $A$ and $U$.
For each word $w$, we perform a $t$-test on the mean difference
between $w_A = \{I(w\in T) ~|~ T\in A\}$ and $w_U = \{I(w\in T) ~|~
T\in U\}$ where $I$ is an indicator function. Table
\ref{table:attr_words} shows some examples of attractive words
identified by our test with p-value $\leq 0.05$. Intuitively, a
title with these words can attract users' clicks for certain
information needs.

\begin{table}[t]\centering
\begin{tabular}{|l|l|} \hline
Category & Attractive Words\\
\hline \hline
Recency & latest, breaking \\\hline
Importance & official, standard, homepage\\\hline
Popularity & images, pictures, video, gallery\\\hline
Others & free, sale, specials, welcome, login\\
\hline\end{tabular}\caption{Sample of attractive words identified
by our $t$-test. Categories are manually labelled.}
\label{table:attr_words}
\end{table}

\subsubsection{URL Attractiveness.} URLs in snippets are also used by end
users to select search results since URLs can implicitly tell users
the reputation or quality of the landing pages \cite{Richardson07}.
For example, a URL with ``.edu'' in its domain is a good indicator
for academic-related queries. A long URL with high depth is
probably less attractive than a URL with low depth if a user
intends to find some broad information. We thus define the URL
attractiveness features as shown in
Table~\ref{tab:snippet-features}. All the URL features are query
independent. For example, although a URL may not received any
clicks for a tail query but it can be probably clicked again if it
has received many clicks in the search logs. This is captured by
our NumViews feature for URLs. Furthermore, we define a categorical
feature TopLevelDomain in Table~\ref{tab:snippet-features} to
roughly capture the URL types. Table~\ref{table:TLD} lists the
distribution of the highly clicked top-level domains identified in
our search logs. The feature TopLevelDomain takes one of the 5
possible values in Table~\ref{table:TLD}.

\begin{table}[t]
\centering
\begin{tabular}{|c||c|c|c|c|c|} \hline
TopLevelDomain & com & org & net & edu & others\\\hline
Percentage & 75.26 & 10.19 & 4.08 & 1.89 & 8.58\\
\hline\end{tabular}\caption{Distribution of top level domain of
highly clicked URLs in our click logs.} \label{table:TLD}
\end{table}

\subsubsection{Matching Attractiveness.}
Query-biased snippets are regarded as the most relevant part of the
landing page by snippet generation methods~\cite{TomSan98}. The
matching fragments of a title, URL and abstract provides passage
level relevance evidence between query and
documents~\cite{Callan:1994} and also play an important role in
users' evaluation of the relevance of the landing page. We define a
set of matching attractiveness in Table~\ref{tab:snippet-features}
in a similar way to the matching features between queries and whole
documents. Our matching features cover string-level match,
token-level match, matching positions (NumBtwMatch and NumBefMatch)
and matching coherence and proximity (IsSegMatch and NumBtwMatch),
etc. We also include approximate matches which are computed based
on the edit distance between query tokens and words in snippets.
This feature can capture the morphological variants and also
acronyms. We discretize the approximate match in to binary values
by thresholding. For example, FracApxMatch is computed as the
fraction of query tokens which have approximate matches in titles
or URLs.

\begin{align*}
\textrm{FracApxMatch}: \frac{1}{|Q|}\sum_{q\in Q}\textrm{ApxMatch}(q,T\cup U)
\end{align*}
where $Q,T,U$ are a set of tokens in the query, title and URL
respectively and
\begin{align*}
\textrm{ApxMatch}(q,S)=\left\{ \begin{array}{ll}
1 & \textrm{if } q \textrm{ approximately matches}\\
& \textrm{a token in } S\\
0 & \textrm{otherwise.}\end{array}\right.
\end{align*}
The longer the token $q$, the more distance we allow in approximate
matches. This feature has been shown to be an important feature in
terms of discriminative power in prediction in our experiments.

Our matching features can be also extended to an expanded set of
queries for a given URL. Though we have no click information for
tail queries, we still have clicked information for a candidate
URL. In our log, we can have a set of queries which have led clicks
to the URL as the expanded set of queries. We can thus compute the
matching attractiveness of this set of queries and use them as
additional snippet features.
Let $Q_{exp}(u)$ denote the set of queries for which the URL $u$
has been viewed and clicked by users in our logs. Given a query $q$
and a URL $u$, we define
\begin{align*}
&\textrm{FracMatch\_Expanded}: \textrm{FracMatch}(q_i, Q_{exp}(u))
\end{align*}
where FracMatch denotes the fraction of query tokens in the
expanded query set.
For example, given $q$=``puma concolor,'' the following URL:\\
\\
URL: en.wikipedia.org/wiki/Mountain\_lion \\
Title: Cougar -Wikipedia, the free encyclopedia\\
\\
We have the expanded query set as \{cougar, mountain lion,
concolor\}. Although there is no matching between the original
query $q$ and the corresponding URL, i.e, FracMatch($q$, $u$) = 0,
we have FracMatch\_Expanded($q$, $u$) = 0.5. This makes sense
because concolor is also known as cougar or mountain lion,
depending on regions. This example shows that expanded query match
features can deal with some synonym or misspelling problems
effectively.

\section{Leverage Snippet Features}\label{sec:strategy}
Given a query, we use $\mathbf{x}_i\in \mathbb{R}^d$ and
$\mathbf{s}_i \in \mathbb{R}^l$ to represent the original ranking
features and the snippet features for document $i$. A traditional
ranking function $f_{org}: \mathbb{R}^d\rightarrow \mathbb{R}$ maps
the original ranking features to a real value and all the documents
for a query is ranked by $f_{org}$ in descent order. We leverage
the snippet features to predict perceived relevance and enhance the
search result ranking.

\subsection{Predict Perceived Relevance} We train an
attractiveness function $f_{attr}:\mathbb{R}^l \rightarrow
\mathbb{R}$ based on the snippet features $\mathbf{s}_i$ and the
attractiveness score $a_i$ estimated using DBN model. We obtain our
training data by applying our feature definition and the DBN model
on a set of popular queries with sufficient click information.
Since $f_{attr}$ only relies on a set of snippet features, it can
be applied to tail queries. We use the GBRank~\cite{ZheZhaZha07}
method to find the optimal $f_{attr}$ to minimize the following
pairwise loss function. Let
$\mathcal{P}=\{(\mathbf{s}_{i},\mathbf{s}_{j},a_{i}-a_{j})\}$. Our
loss function is:
\begin{align*}
\sum_{\mathcal{P}}\max \Big(0,(a_{i}-a_{j})-(f_{attr}(\mathbf{s}_{i})-f_{attr}(\mathbf{s}_{j}))\Big)^2.
\end{align*}
The function $f_{attr}$ can be used to predict the perceived
relevance between any query and URL.

\subsection{Improve Web Search Ranking}
In this section, we discuss how to leverage our snippet features to
enhance the ranking. We propose two strategies and discuss their
application scenarios in the following.

\subsubsection{Strategy I}
To leverage the snippet features, our first strategy is to use the
predicted perceived relevance scores and combine them with the
original ranking scores to rerank the top search results.
Specifically, we propose the following scenario to apply our
strategy:

\begin{itemize}
\item An initial query is issued and the ranking function
    $f_{org}$ is used to select a few top results.
\item The snippet generation method receives the selected
    documents. It generates the snippets and also the snippet
    features. Based on snippet features, $f_{attr}$ is used to
    estimate the perceived relevance.
\item The final ranking of search results is ranked based on a
    linear combination of $f_{org}$ and $f_{attr}$:
    $$f_{I} = \lambda \cdot f_{org}+(1-\lambda)\cdot f_{attr}.$$
\end{itemize}

\subsubsection{Strategy II}
The first strategy is a simple linear combination of the predicted
scores. Our second strategy is to go to the feature level and
expand the ranking features $\mathbf{x}_i$ by $\mathbf{s}_i$. Thus
we form a longer feature vector $[\mathbf{x}_i, \mathbf{s}_i]$ for
each document. We train a new ranking function $f_{I}:
\mathbb{R}^{d+l} \rightarrow \mathbb{R}$ on these concatenated
vectors. Apparently, it is difficult to directly apply such a
strategy on a search engine since the search snippets features can
be generated only after the snippets are generated. We thus propose
the following scenario in a feedback setting to have two rounds of
retrieval.

\begin{itemize}
\item An initial query is issued and the ranking function
    $f_{org}$ is used to return the search results and generate
    snippets for the top ranked results.
\item We provide users an additional button ``Refresh to
    Improve'' which is intended to improve search results if a
    user is not satisfied with the current results and clicks
    the button.
\item After the button is pressed, all the snippet features are
    generated for top results and the new ranking function
    $f_{II}$ is used to generate a new search result page.
\end{itemize}

This strategy can be used without user interference by search
engines. However, such a strategy may be risky for those queries
for which the original ranking is already good enough. The button
``Refresh to Improve'' is a safe alternative when a user is not
satisfied with the current results.

\section{Experiments}\label{sec:exp}

We perform two types of experiments. First, we evaluate the
performance of our attractiveness prediction. Then, we use the
predicted attractiveness and the defined snippet features to
improve the ranking accuracy.

\subsection{Predict Perceived Relevance}

\subsubsection{Experiment Setup}
We first test the predictive accuracy of our proposed prediction
model for tail queries. A difficulty in this test is that we cannot
obtain the ``true" target attractiveness for tail queries: The
estimation of attractiveness (by click models) is not reliable for
tail queries due to the limited amount of click information. Thus,
we need to \emph{simulate} tail queries by sampling only a small
subset of click logs of non-tail queries. Please note that the
target attractiveness is obtained before the sampling.

We get click logs from  a commercial search engine. The click log
data is a set of sessions. A session is associated with a unique
user and a unique query. It starts when a user issues a query and
ends with 60 minutes idle time on the user side. Each session
contains the list of URLs in the search results page and list of
clicked URLs. We select queries with enough sessions to ensure the
reliable target values. After this filtering, we obtain 40M
sessions and 20K unique queries. Let this original set of sessions
be $S$. Then, $S$ is split into the training set $S_{train}$, the
validation set $S_{validation}$ and the test set $S_{test}$. For
all (query,URL) pairs in these sets, we obtain snippet features and
target values (attractiveness computed by the DBN click model using
the full data). We get $S^{tail}_{test}$ by sampling 10 random
sessions for each query in $S_{test}$.

The evaluation is based on comparing pairwise attractiveness values
predicted by our proposed model $f_{attr}$ to the ``true" pairwise
attractiveness values derived from the DBN click model using the
full session data: For two URLs, $u_j$ and $u_k$ for query $i$, we
predict that URL $u_j$ is more attractive than URL $u_k$ if
$$f_{attr}(\mathbf{x}_{i,j})- f_{attr}(\mathbf{x}_{i,k})>\tau.$$
Then, we test if $a_{i,j}>a_{i,k}$ where $a_{i,j}$ and $a_{i,k}$
are the true target attractiveness values computed by the DBN click
model using the full session data $S_{test}$. Hence, this is a
binary classification problem. With a different $\tau$ values, we
have a different levels of precision and recall. Thus, by varying
$\tau$, we can get a precision-recall curve.

Based on the test data, we compare the predictions given by the
following:
\begin{itemize}
\item $a^{tail}$ : Attractiveness computed by the DBN click
    model using sampled session data $S^{tail}_{test}$.
\item $f_{snippet}$ : Function trained on only snippet
    features.
\item $f_{snippet+click}$ : Function trained on both snippet
    features and clicks.
\end{itemize}

$a^{tail}$ provides the baseline predictions: If $a^{tail}_{i,j}$
$-$ $a^{tail}_{i,k}$ $>\tau$, we predict that URL $u_j$ is more
attractive than URL $u_k$.

We apply GBRank to train a function $f_{snippet}$ on the pairwise
training data
\begin{align*}\mathcal{P}=
&\{(\mathbf{x}_{i,j},\mathbf{x}_{i,k},a_{i,j}-a_{i,k})| \\
&i\in \{1,\ldots,N\},j,k\in \{1,\ldots,10\}, a_{i,j}>a_{i,k}\}
\end{align*}
where $a_{i,j}$ and $a_{i,k}$ are the attractiveness values
computed by the DBN click model using the whole session data
$S_{train}$.

Once we train $f_{snippet}$ on the training data, it can be used
for new queries for which no click information is available.
However, for tail queries for which some amount of click data is
available, we can combine our attractiveness model and the click
information. A straightforward way to combine the two is to have a
linear combination of the two predictions:
\begin{align*}
\lambda f_{snippet}+(1-\lambda)a
\end{align*}
where $a$ is the attractiveness computed by the DBN click model
using the available click data and $\lambda$ depends on the
frequency of a query (the more frequent, the smaller $\lambda$
becomes). However, we would have to tune $\lambda$ manually or
design a heuristic function for $\lambda$. More principled way of
combining the attractiveness model and click information is to use
the click information as a feature and let the training procedure
figure out the optimal combination. To this end, we generate
another session data $S'_{train}$ as follows. For each query in
$S_{train}$, we sample $r$\% of sessions where $r$ is randomly
selected to ensure that the sampled session data contains queries
with various frequencies. Then, each feature vector
$\mathbf{x}_{i,j}$ in our training data is expanded to include two
additional features:
\begin{itemize}
\item $a'_{i,j}$ : Attractiveness computed by the DBN click
    model using $S'_{train}$
\item $session_i$ :  The number of sessions for query $i$ in
    $S'_{train}$
\end{itemize}
Note that we still use the true attractiveness $a_{i,j}$ (computed
by using the full data $S_{train}$) as targets. The new pairwise
training data is
\begin{align*}
\mathcal{P}&=\{((\mathbf{x}_{i,j},a'_{i,j},session_i),(\mathbf{x}_{i,k},a'_{i,k},session_i),a_{i,j}-a_{i,k})\\
&|~ i\in \{1,\ldots,N\}, j,k\in \{1,\ldots,10\}, a_{i,j}>a_{i,k}\}.
\end{align*}
Then, we train a GBrank function $f_{snippet+click}$ on this data.

\subsubsection{Experimental Results}\label{sec:ex_predict_attr}
\begin{figure}
\begin{center}
    \includegraphics[width=80mm]{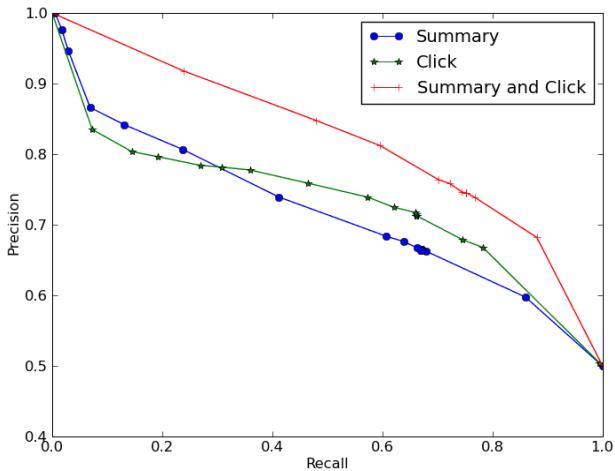}
\end{center}
\caption{Precision vs. recall of 3 different ways of predicting
attractiveness for tail queries. `Summary' represents a function $f_{snippet}$
trained on only snippet features. `Click' represents predictions given by $a^{tail}$,
attractiveness computed by the DBN click model using a limited amount of
click information. `Summary and Click' represents a function
$f_{snippet+click}$ that combines both predictions.}
\label{fig:1}
\end{figure}

We summarize the precision-recall results of $a^{tail}$,
$f_{snippet}$ and $f_{snippet+click}$ in Figure \ref{fig:1}. The
result shows that the combination of our attractiveness model and
the click information clearly outperforms either one.

After the training process, we obtain the list of features ordered
by their importance (See \cite{Fri01} for the definition of
importance of features). We have the following observations:
\begin{itemize}
\item For $f_{snippet+click}$, the attractiveness and the
    number of sessions are among top three features in the
    importance list. When we look into the decision tree
    structure, we find that the two features function together:
    When the number of sessions is large (i.e. we have
    sufficient click information), the attractiveness computed
    by the DBN click model should be weighted more than snippet
    features. On the other hand, when we have a small number of
    sessions (i.e. tail queries), snippet features should play
    a more important role.
\item Length of URL is the second most important feature for
    $f_{snippet}$ and the fourth for $f_{snippet+click}$, which
    agrees with the results by \cite{clarke2007click}.
\item Features related to URL and title are more important than
    those for abstract.
\end{itemize}

\subsection{Improve Ranking Relevance}

We construct our data sets to test the effectiveness of our defined
snippet features from a commercial search engine. The training
examples are labeled using five values, $\{0,1,2,3,4\}$,
representing five levels of relevance. Our evaluation is based on
NDCG$_5$ and NDCG$_1$. NDCG$_k$ is defined to be
\begin{align*}
\textrm{NDCG}_k=\frac{1}{Z_k}\sum_{i=1}^k\frac{G_i}{\log_2(i+1)}
\end{align*}
where $G_i$ is the function of relevance grade of the document at
rank position $i$ and $Z_k$ represents a normalization factor to
guarantee that the NDCG$_k$ for the perfect ranking (among the
permutations of the retrieved documents) is 1.

We have a conventional data set which has the most informative 20
original ranking features, including some click-based features, to
train a conventional ranking function.
Since we aim at improving relevance for new or tail queries, we
collect (query,URL) pairs which have no click related information
from the above data set. We treat all the queries in the resulting
data set as tail queries. Table \ref{table:dcg_tail_query_length}
shows the distribution of the tail queries with respect to query
length and their corresponding search accuracy using our baseline
ranking function. Clearly, long queries cover a large portion of
the tail queries. Furthermore, we can also see that while short
tail queries can achieve reasonable accuracy, long queries usually
have much worse search accuracy. This means that the baseline
ranking function is less effective for longer tail queries. Thus,
in the following experiments, we consider the queries with more
than or equal to 3 tokens to help these more difficult tail
queries. We split the data into training and test. In the training
data, we have 202K (query,URL) pairs, resulting in 2M preference
pairs. In the test data, we have 46K (query,URL) pairs and 545K
preference pairs. Since no click information is available, all the
queries in both training and test data can be regarded as unseen
queries.

\begin{table}
\centering
\begin{tabular}{|c|c|c|}\hline
& Ratio of tail queries & NDCG$_5$ \\\hline \hline
1 token & 8.55\% & 0.801 \\\hline
2 token & 18.0\% & 0.720 \\\hline
3 token & 32.5\% & 0.653 \\\hline
+4 token & 46.3\% & 0.588 \\
\hline\end{tabular} \caption{Ratio of tail queries and search
quality broke down by query length.}
\label{table:dcg_tail_query_length}
\end{table}

To obtain the training data to learn the attractiveness function
$f_{attr}$, we use the data set used in the previous section. Each
session contains the list of URLs in the search result page and
list of clicked URLs. We select queries with enough sessions to
ensure the reliable target values. 

\begin{figure}[t]
\centering
\includegraphics[width=80mm]{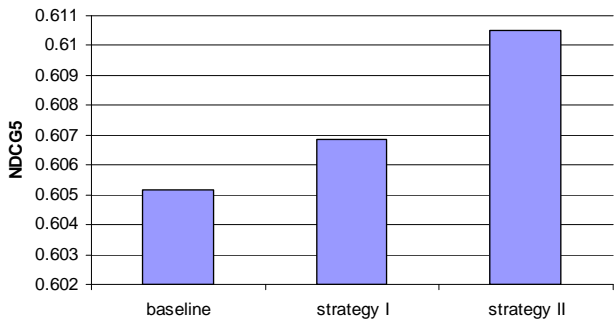}
\caption{Comparison of NDCG$_5$ for the baseline function $f_{org}$
the strategy I ranking function $f_{I}$ and the strategy II method $f_{II}$.}
\label{fig:maincmp}
\end{figure}

Figure~\ref{fig:maincmp} shows the accuracy comparison of the
baseline ranking ($f_{org}$), strategy I ($f_{I}$), and strategy II
($f_{II}$) using NDCG$_5$ as the metric. For all these methods, we
tune the GBRank parameters and $\lambda$ to be the optimal. From
this figure, we can see that both our strategies can improve over
the baseline ranking. For example, strategy II improve over the
baseline by $0.8\%$ relatively and this is statistically
significant based on the Wilcoxon test (p-value $<$ 0.01). Although
strategy I is also able to improve over the baseline, the
improvement is not statistically significant. Comparing the two
strategies, strategy II is more effective than strategy I. This
shows that the second strategy of directly training a new ranking
function can better leverage the snippet feature signals.

\begin{figure}[t]
\centering
\includegraphics[width=80mm]{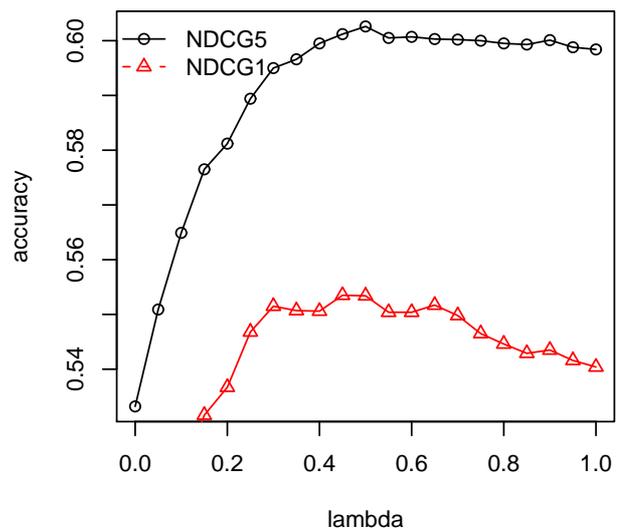}
\caption{The impact of the linear combination factor $\lambda$ of strategy I.}
\label{fig:lambda}
\end{figure}

In Figure~\ref{fig:lambda}, we show the impact of the parameter
$\lambda$ in strategy I using both metrics NDCG$_5$ and NDCG$_1$.
When $\lambda=1$, the result is the same as the baseline. From this
figure, we can see that strategy I can only marginally improve over
the baseline method in terms of NDCG$_5$. However, we observe
significant improvement of strategy II over baseline in terms of
NDCG$_1$. For example, when $\lambda=0.5$, the NDCG$_1$ of $f_{I}$
is 0.554 and achieves $2.6\%$ relative improvement over 0.540 of
$f_{org}$ and this improvement is also statistically significant.
This means that our strategy I is more effective for higher ranked
documents. This also means that the attractiveness scores from DBN
is more accurate to predict higher ranked results and this is
reasonable because the highly ranked documents is less influenced
by the position bias.

Overall, we can see that both our strategies are effective to
improve search accuracy. This confirm the effectiveness of our
defined snippet features.

\section{Conclusions and Future Work}\label{sec:con}
In this paper, we studied how to model perceived relevance for
tails queries without relying on any click-through data. We
developed a set of snippet features to capture the attractiveness
or perceived relevance of Web search results and proposed two novel
strategies to leverage these snippet features to improve tail
queries. We show that the two strategies can be naturally
incorporated into a search process. We conduct experiments on a
large data set from a commercial search engine. Our results confirm
the defined snippet features are able to predict the perceived
relevance effectively. Furthermore, the search accuracy of tail
queries can be significantly improved by using the snippet
features.

Our work is one of the few work on directly improving search
accuracy for tail queries. In the future, one interesting direction
is to provide a unified framework to jointly model both clicks and
snippet features together so that information of head queries can
be propagated to tail queries in a more principled way. A main
challenge for tail queries is due to lack of users' feedback and a
possible direction is to leverage the relation between queries such
as a query graph to better capture the attractiveness of search
results for tail queries.

\bibliographystyle{plain}
\bibliography{clks}

\end{document}